\documentclass[useAMS,usenatbib,usegraphicx]{mn2e}

\usepackage{graphicx}
\usepackage{txfonts}
\usepackage{rotating}
\usepackage{array}
\usepackage{natbib}

\def\hen{Hen\,3-209}
\def\l{$\lambda$}
\def\ha{H$\alpha$}
\def\hb{H$\beta$}

\def\he{He\,{\sc i}}

\def\nii{[N\,{\sc{ii}}]}

\def\feii{Fe\,{\sc{ii}}}
\def\sii{[S\,{\sc{ii}}]}
\def\oiii{[O\,{\sc{iii}}]}
\def\kms{km\,s$^{-1}$}

\def\ros{\emph{{\sc ROSAT}}}

\title[The atypical emission-line star Hen\,3-209]{The atypical emission-line star Hen\,3-209\thanks{Based on observations collected at the European Southern Observatory (La Silla, Chile) and the Cerro-Tololo Observatory (Chile).}}

\author[Y. Naz\'e et al.]{Y.\ Naz\'e$^1$\thanks{Postdoctoral Researcher FNRS (Belgium)}\thanks{E-mail: naze@astro.ulg.ac.be}, G.\ Rauw$^1$\thanks{Research Associate FNRS (Belgium)}, D.\ Hutsem\'ekers$^{1}$\footnotemark[4], E.\ Gosset$^{1}$\footnotemark[4],  J.\ Manfroid$^1$\thanks{Research Director FNRS (Belgium)},  P.\ Royer$^2$ \\
$^1$ Institut d'Astrophysique et de G\'eophysique, Universit\'e de Li\`ege, All\'ee du 6 Ao\^ut 17, B\^at B5c, 4000 Li\`ege, Belgium \\
$^2$ Instituut voor Sterrenkunde, Katholieke Universiteit Leuven, Celestijnenlaan 200B, 3001 Leuven, Belgium}

\begin{document}


\maketitle

\label{firstpage}

\begin{abstract}
We analyse observations, spanning 15 years, dedicated to the extreme emission-line object \hen. Our photometric data indicate that the luminosity of the star undergoes marked variations with a peak-to-peak amplitude of 0.65~mag. These variations are recurrent, with a period of 16.093$\pm$0.005~d. The spectrum of \hen\ is peculiar with many different lines (H\,{\sc{i}}, \he, \feii,...) showing P Cygni profiles. The line profiles are apparently changing in harmony with the photometry. The spectrum also contains \oiii\ lines that display a saddle profile topped by three peaks, with a maximum separation of about 600~\kms. \hen\ is most likely an evolved luminous object suffering from mass ejection events and maybe belonging to a binary system. 
\end{abstract}

\begin{keywords}
stars: emission-line -- stars: individual: Hen\,3-209 -- stars: peculiar
\end{keywords}

\section{Introduction}

In the middle of the last century, dedicated \ha\ surveys have identified many emission-line stars. A large number of these stars were at first classified as Wolf-Rayet (WR) stars but, on second sight, it appears that their spectrum rarely presents the typical WR characteristics. Although these peculiar objects may represent keys for understanding stellar evolution, only a few of them, generally the brightest and least reddened ones, were analysed in detail. 

\hen, also known as WRAY\,15-285, MR\,14, SS73\,10, or Ve\,6-14 ($8^{\rm h}48^{\rm m}45.5^{\rm s} -46 ^{\circ}05'09''$, J2000), is such an emission-line object \citep{vel}. It has only been poorly studied since its detection despite its peculiarities. The star actually displays a strong and broad \ha\ line which led to the classification of \hen\ as a Wolf-Rayet star \citep{roberts,henize}, an extreme Be-like object \citep{sand} and more recently as a B[e] star \citep{dew}. \citet{henize} even suggested that \hen\ could be a nova on the basis of its non-detection by \citet{smi}. However, as the star is photometrically variable (see below), it could simply have been below Smith's detection limit at the time of her observation.

The aim of this paper is to present the results of a long-term observing campaign on this object. The paper is organized as follows: the observations are presented in Sect.\,2, the photometric data are analysed in Sect.\,3, and the spectral features of \hen\ are examined in Sect.\,4. Finally, we discuss in Sect.\,5 the nature of the star using the gathered evidence and we conclude in Sect.\,6.

\section{Observations and data reduction}
\subsection{Photometry\label{phot}}
Between 16 March and 19 April 1997, we performed differential photometry of \hen\ with the 0.6\,m Bochum telescope at La Silla, Chile, equipped with a direct camera and a Thomson 7882 CCD detector subtending a field of 3.2\arcmin\ $\times$ 4.8\arcmin. All the observations have been performed through a Johnson $V$ filter (exp. time 5 min). About 60 independent science frames have been acquired distributed all over the run. Details on the particular treatment of flat-field calibrations can be found in \citet{GOS}. 

On each reduced frame, we performed aperture photometry of all the objects down to some threshold. The field of \hen\ is sufficiently populated so that we can use other stars on the same CCD frame as constant reference stars. The photometry (Table~\ref{tab: obsp}) was further reduced in the standard way through a global minimization process allowing for extinction, zero point of individual frames, etc. The typical dispersion in the final relative magnitudes of a constant star with the same brightness as \hen\ is characterized by $\sigma=0.01$~mag. On a few nights, we also observed standard stars in order to perform absolute photometry. The latter procedure allowed us to fix the zero point magnitude with a typical error of $\sigma=0.04$~mag.

\begin{table}
\centering
\caption{Photometric observations of \hen. Date is the Heliocentric Julian Date (in the format HJD$ - 2\,400\,000$) at mid-exposure, while $V$ corresponds to the differential $V$ magnitude measured with an aperture of 2\arcsec. 
\label{tab: obsp} }
\begin{tabular}{c c  | c c | c c}
\hline
\vspace*{-0.3cm}&&\\
HJD & $V$  & HJD & $V$ & HJD & $V$ \\
\hline
50528.5809 & 13.574 & 50537.5412 & 13.652 & 50547.5019 & 13.547\\
50528.5821 & 13.558 & 50538.5283 & 13.449 & 50547.5902 & 13.530\\
50528.5892 & 13.581 & 50538.5722 & 13.444 & 50548.5092 & 13.516\\
50528.5931 & 13.581 & 50538.6109 & 13.453 & 50548.5612 & 13.519\\
50528.6710 & 13.585 & 50538.6326 & 13.462 & 50548.6170 & 13.522\\
50528.6753 & 13.575 & 50539.5336 & 13.392 & 50549.5526 & 13.570\\
50528.6788 & 13.587 & 50539.5765 & 13.394 & 50550.5540 & 13.678\\
50530.5933 & 13.539 & 50539.6396 & 13.369 & 50550.6023 & 13.685\\
50530.7505 & 13.531 & 50540.5393 & 13.310 & 50551.5890 & 13.878\\
50531.5381 & 13.552 & 50540.5755 & 13.299 & 50552.5046 & 13.921\\
50531.6509 & 13.548 & 50540.6194 & 13.305 & 50552.6094 & 13.915\\
50532.5553 & 13.517 & 50541.5073 & 13.342 & 50553.5593 & 13.730\\
50532.6390 & 13.517 & 50541.6484 & 13.358 & 50554.5574 & 13.556\\
50533.5856 & 13.607 & 50542.5106 & 13.373 & 50556.5014 & 13.305\\
50533.6404 & 13.608 & 50542.5469 & 13.357 & 50556.5685 & 13.306\\
50534.5449 & 13.772 & 50543.5067 & 13.470 & 50557.5083 & 13.325\\
50534.5976 & 13.765 & 50543.5610 & 13.465 & 50557.5631 & 13.323\\
50534.6436 & 13.773 & 50544.4993 & 13.572 & 50557.5932 & 13.328\\
50535.5349 & 13.901 & 50544.6015 & 13.555 & 50558.5039 & 13.392\\
50535.5767 & 13.899 & 50545.5918 & 13.506 & 50558.5513 & 13.406\\
50535.6171 & 13.879 & 50546.4946 & 13.555 & \\
50535.6385 & 13.865 & 50546.5739 & 13.527 & \\
\hline
\end{tabular}
\end{table}

\begin{table}
\centering
\caption{Journal of the spectroscopic observations of \hen. Column 1, 2, 3, 4  and 5 indicate respectively the year of the observation, the instrumental configuration used, the phase (see Sect. 3), the Heliocentric Julian Date (in the format HJD $- 2\,400\,000$) at mid-exposure and the wavelength range covered (in \AA). The typical exposure times (in minutes) is shown in the last column.
\label{tab: obs} }
\begin{tabular}{l l c c cc}
\hline
\vspace*{-0.3cm}&&\\
Date& \multicolumn{1}{c}{Instrument} & $\phi$& HJD & Range & $T_{exp}$ \\
\hline
11/90   & ESO~1.5\,m + B\&C & .41&48225.675 &[4550-&[10-\\
        & RCA CCD (ESO 13)& .41 & 48225.694 &-7400]&-50]\\
        &grating 25 (Res. 6.5\AA)& .42 & 48225.721 & \\
        & 1'' slit&.47& 48226.647 & \\
12/90   &grating 23 (Res. 4.5\AA)&.54& 48227.682 &[3600-& 60 \\
        &1.2'' slit&.54 &48227.729 &-5500]\\
04/91   & ESO~1.5\,m + B\&C &.42&48370.673 &[4300-& 30 \\
        &Thomson 1K CCD&.42&48370.693 &-4950]\\
        &  (ESO 18), 1.4'' slit&.47&48371.488 \\
        &grating 20 (Res. 1.4\AA)&.48&48371.510 \\
        &&.48&48371.533 \\
        &&.48&48371.556 \\
        &&.54&48372.486 \\
        &&.54&48372.509 \\
04/96   & ESO~1.5\,m + B\&C &.89&50180.553 &[3800-& 30 \\
        &thinned UV flooded&.95&50181.600 &-4800]\\
        &CCD (ESO 39), 2'' slit&.01&50182.532 \\
        &grating 32 (Res. 1.2\AA)&.01&50182.587 \\
04/96   & CAT + CES  & .31&50203.519 &[6520-& $<$60 \\
        &Loral 2688$\times$512 CCD&.31&50203.546 &-6610]\\
        &(ESO 38), 2'' slit,&.39&50204.664 \\
        &Res. 0.1\AA&.39&50204.672 \\
03/97   & ESO~1.5\,m + B\&C &.89&50534.592 &[3860-&30\\
        &same as Apr. 1996&.95&50535.603 &-4830]\\
        &&.01&50536.591 \\
        &&.07&50537.587 \\
        &&.13&50538.579 \\
        &&.20&50539.605 \\
05/99   & CTIO~1.5\,m + Cspec&.91&51323.453 &[4277-& 45 \\
        & Loral 1200$\times$800 CCD & &&-4925]\\
        & \multicolumn{2}{l}{grating 47 (Res. 1.6\AA),2'' slit}\\
03/02   & NTT + EMMI (REMD) &.92&52353.669 &[4040-&60\\
        & grat. 9+grism 3,&.97&52354.515 &-7670]\\
        & Res. 0.8\AA, 1'' slit\\
09/05   & ESO~3.6\,m + EFOSC2 & .78&53638.890 &[6000-& 5\\
        & \multicolumn{2}{l}{thinned Local CCD, 1'' slit} &&-10300]\\
        & grism 16 (Res. 13\AA)\\
\hline
\end{tabular}
\end{table}

\subsection{Spectroscopy}
We have observed \hen\ spectroscopically over a period of 15 years with various instruments (Table \ref{tab: obs}). A low resolution red/near-IR spectrum was obtained in September 2005 during test time on the EFOSC2 instrument attached to the ESO 3.6\,m telescope. Medium resolution spectra were gathered in 1990, 1991, 1996 and 1997 with the ESO 1.5\,m telescope equipped with a Boller \& Chivens (hereafter B\&C) Cassegrain spectrograph. One additional medium resolution spectrum was obtained at the 1.5\,m Ritchey-Chretien telescope of the Cerro Tololo Inter-American Observatory (CTIO) with the Cspec Cassegrain spectrograph in May 1999. High resolution spectra, covering roughly 90\,\AA\ and centered on the \ha\ line, were obtained in 1996 with the 1.4\,m Coud\'e Auxiliary Telescope (CAT) at La Silla, feeding the Coud\'e Echelle Spectrometer (CES) equipped with the Long Camera (LC). Finally, two echelle spectra of \hen\ were gathered in March 2002 with the EMMI instrument at ESO's New Technology Telescope (NTT) at La Silla. 

\section{Photometric variability of \hen}

Two observations taken in February 1991 showed that \hen\ displayed on 24 Feb. $V$=13.48~mag, $V$--$R$=0.92~mag and $V$--$I$=1.76~mag with a $\sigma$ of 0.01~mag, whereas on 28 Feb. the values were $V$=13.90~mag, $V$--$R$=1.03~mag, $V$--$I$=1.83~mag, $B$--$V$=1.04~mag, and $U$--$B$=0.23~mag with a $\sigma$ of 0.05~mag.
\begin{figure*}
\begin{center}
\includegraphics[bb=23 200 580 520, clip, width=8.cm]{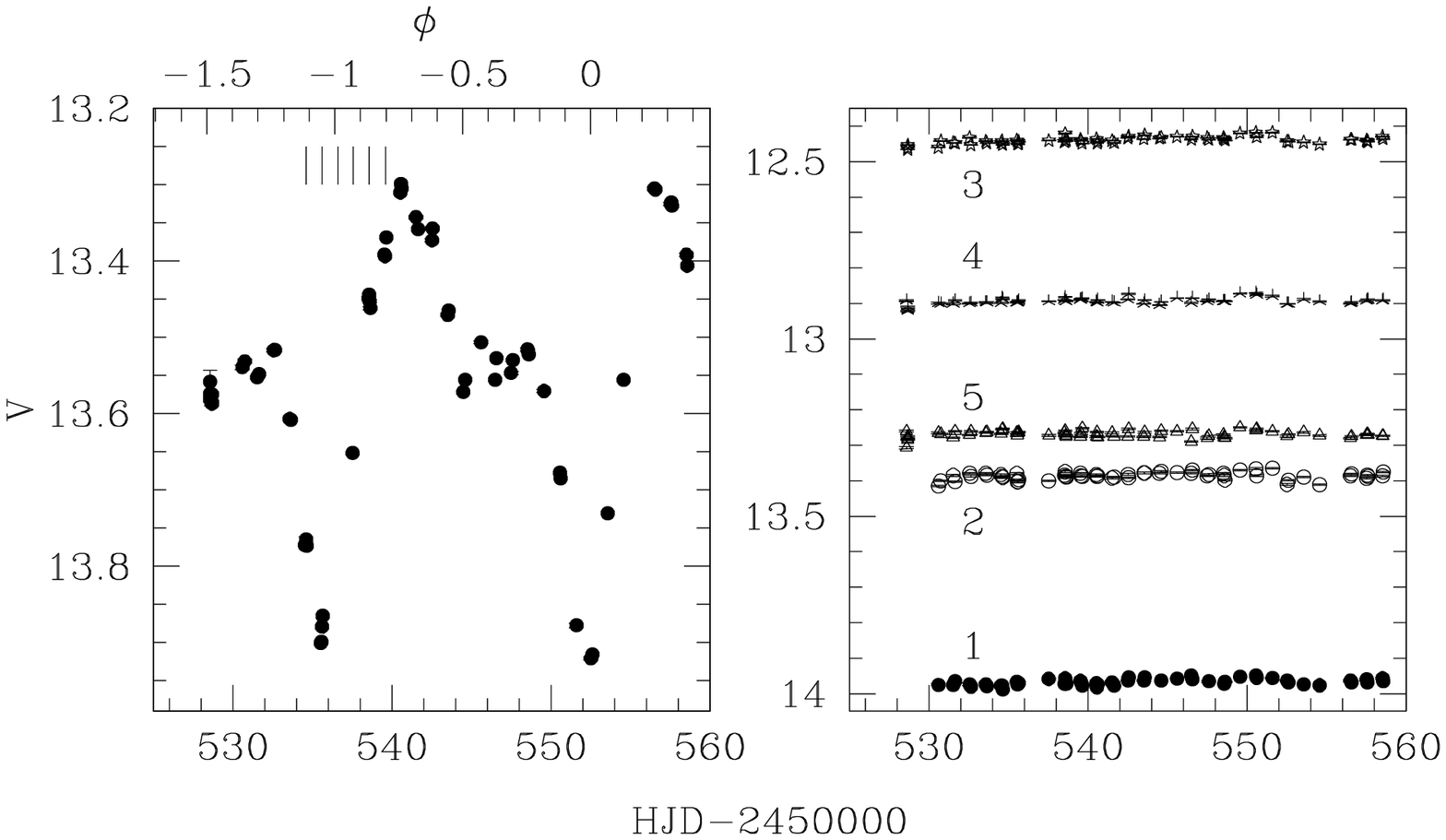}
\includegraphics[width=5.cm]{hen3-209_refphot.ps}
\includegraphics[width=4.5cm]{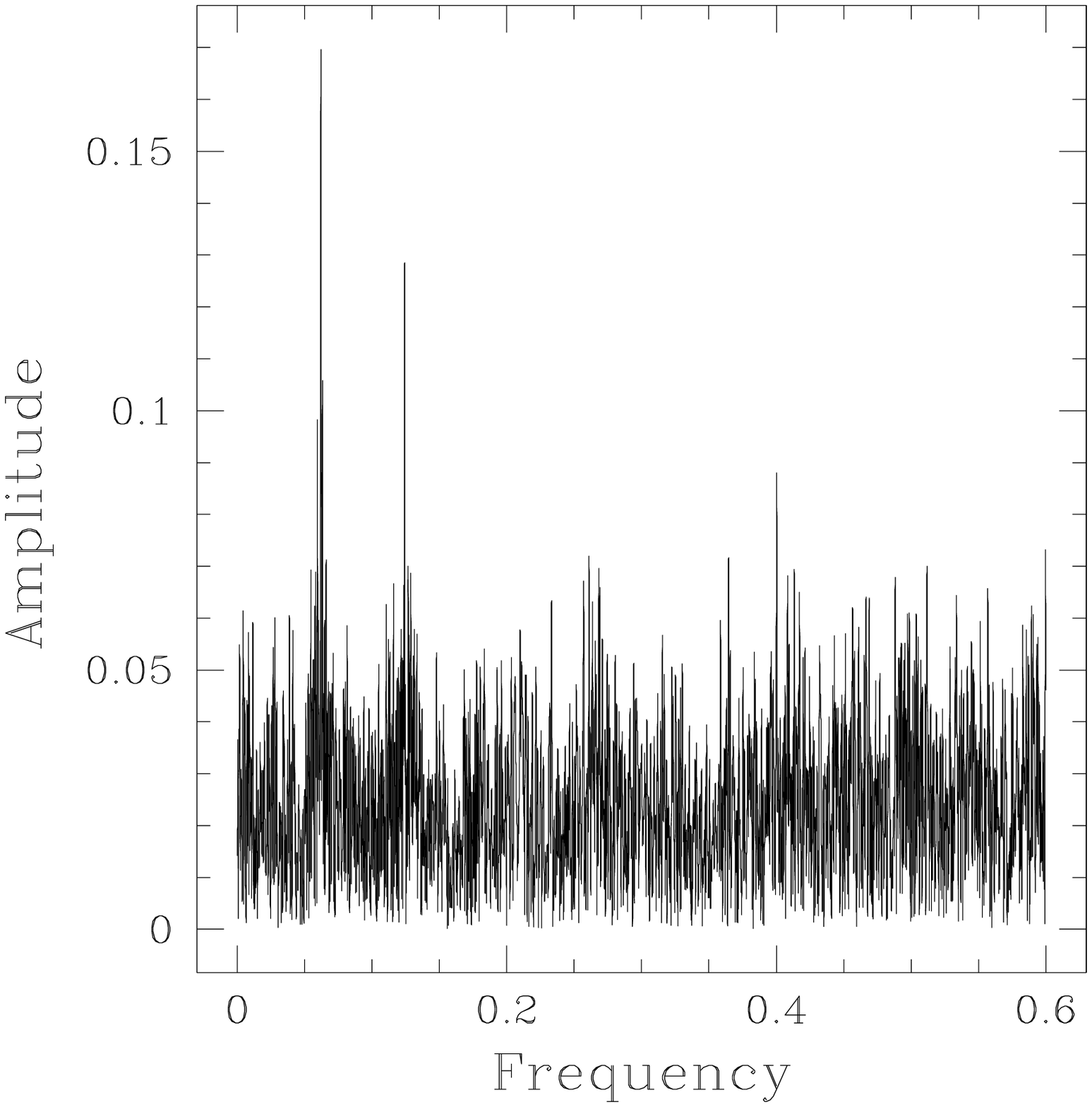}
\caption{\label{fig: phot} $V$-band photometry of \hen\ (first panel) and of the 5 brightest stars in its neighbourhood (second panel). The third panel shows the position of these comparison stars with respect to \hen. The periodogram of the ASAS photometry of \hen\ is displayed in the fourth and last panel: it shows the presence of the 16~d period (corresponding to a frequency of 0.062~d$^{-1}$) and its harmonics. Note that in the first panel, the top vertical ticks indicate when the simultaneous spectroscopic observations were performed (Table \ref{tab: obs}). }
\end{center}
\end{figure*}

Since the luminosity of the star appeared variable, we undertook a monitoring of \hen\ with the Bochum telescope. This photometry (reported in Table \ref{tab: obsp} and shown in Fig. \ref{fig: phot}) confirms the large variations of the luminosity of \hen, with a peak-to-peak amplitude of 0.65~mag and a $\sigma$ of 0.17~mag. In comparison, the 5 brightest stars of the field have constant $V$ magnitudes with $\sigma$ of only 0.01~mag, much lower than the detected variations of \hen\ (Fig. \ref{fig: phot}).

As these variations seem to be recurrent, we have further analysed the data with the method of \citet[ hereafter LK]{LK} and a Fourier-type one (method of \citealt{HMM}, hereafter HMM, see also the comments by \citealt{GOS}). Peaks in the periodogram corresponding to periods $P_{\rm LK}=16.7\pm1.0$~d and $P_{\rm HMM}=17.5\pm1.0$~d were detected. Such periods are confirmed by an auto-correlation analysis, that unambiguously shows the presence of $P\sim17\pm1.0$~d. 

However, only two cycles were measured with the Bochum telescope and this periodicity could thus be spurious. Although no data covering a longer time base are available in the literature, archives from the All-Sky Automated Survey \citep[ASAS,][]{poj}\footnote{Available from http://archive.princeton.edu/ \~\ asas} provided 214 photometric measurements of \hen\ spanning 2000~days, i.e. more than 100 cycles (from Nov. 2000 to Feb. 2006). These data present a much larger dispersion than ours ($\sigma$=0.13~mag for the comparison stars) but they clearly confirm the periodic behaviour of \hen. In addition, since the time base is longer, the determination of the period is much better: $P_{\rm LK}=16.103\pm0.013$~d and $P_{\rm HMM}=16.090\pm0.013$~d (see the periodogram on Fig. \ref{fig: phot}).

Combining the ASAS and Bochum data yields a period of $P=16.093\pm0.005$~d. All the phases quoted in this paper refer to this final value, taking into account an arbitrarily chosen $T_0$ (in HJD) of 2\,450\,552.5046 which corresponds to the minimum light in the Bochum data.

We might note that our average magnitude of \hen\ is in agreement with the recent value $V= 13.56$ of \citet{dew}, but it is however larger than the magnitude of 12.5 given by \citet{sand}. Although this latter value might be more subject to uncertainties, it is also possible that \hen\ was indeed brighter some 30 years ago. 

\begin{figure*}
\begin{center}
\includegraphics[bb=45 145 505 690, clip, width=12cm]{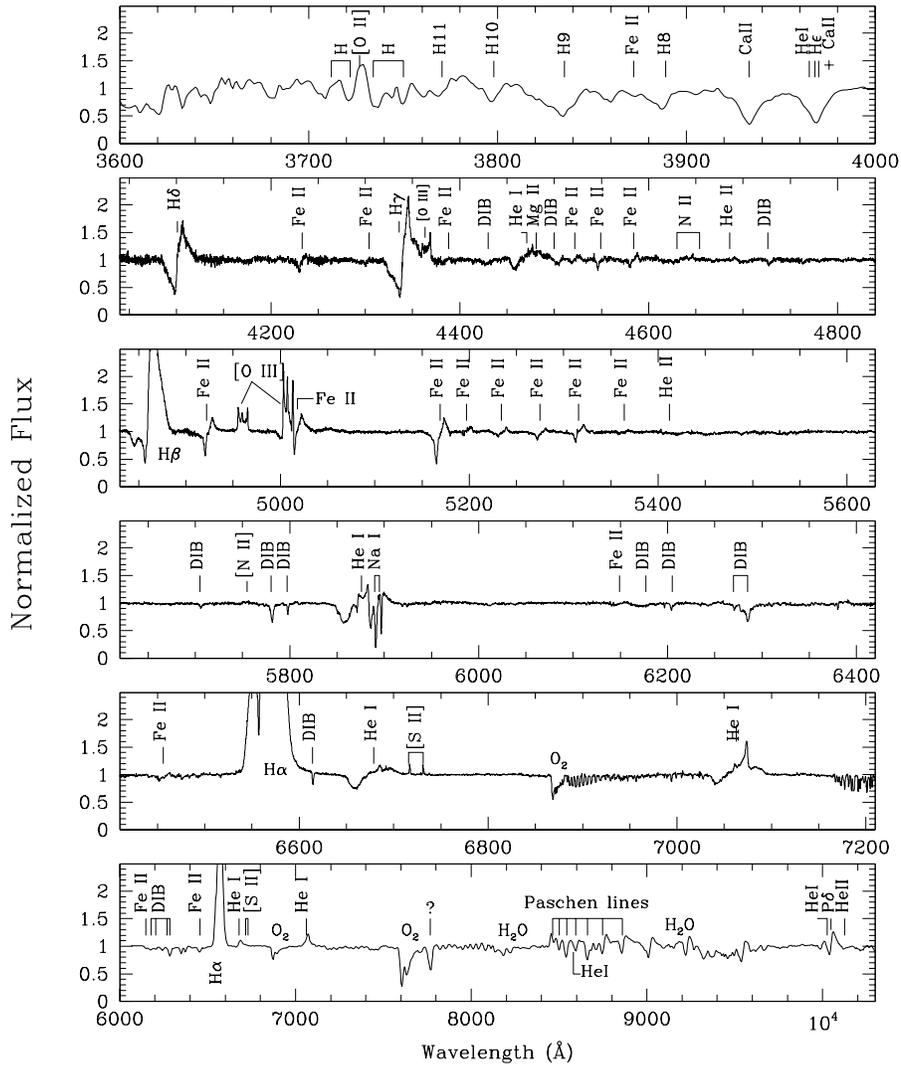}
\caption{\label{echelle} The spectrum of \hen: the top panel shows B\&C data from 1990, the four middle panels present the EMMI echelle spectrum as observed in March 2002, and the bottom panel displays the red/near-IR data obtained in 2005 with EFOSC2. The most important stellar and interstellar lines as well as diffuse interstellar bands (DIBs) are indicated. An unidentified feature appears at 7767\AA; it is not associated to any known line, DIB or instrumental defect.}
\end{center}
\end{figure*}

\section{The spectrum of \hen}
Figure\,\ref{echelle} presents the spectrum of \hen. It is dominated by a huge H$\alpha$ emission and exhibits broad P Cygni profiles in the upper Balmer lines as well as in a number of He\,{\sc i} and Fe\,{\sc ii} transitions. A very weak He\,{\sc ii} $\lambda$\,4686 emission might also be present, whereas the red/near-IR spectrum revealed only a few Paschen and helium lines. The analysis of the most prominent lines of this spectrum generally reveals three main features: a redshifted emission component plus two blueshifted absorptions, a narrow one and a broad one. 

\begin{figure*}
\begin{center}
\includegraphics[bb=30 170 400 510, clip, width=4.2cm]{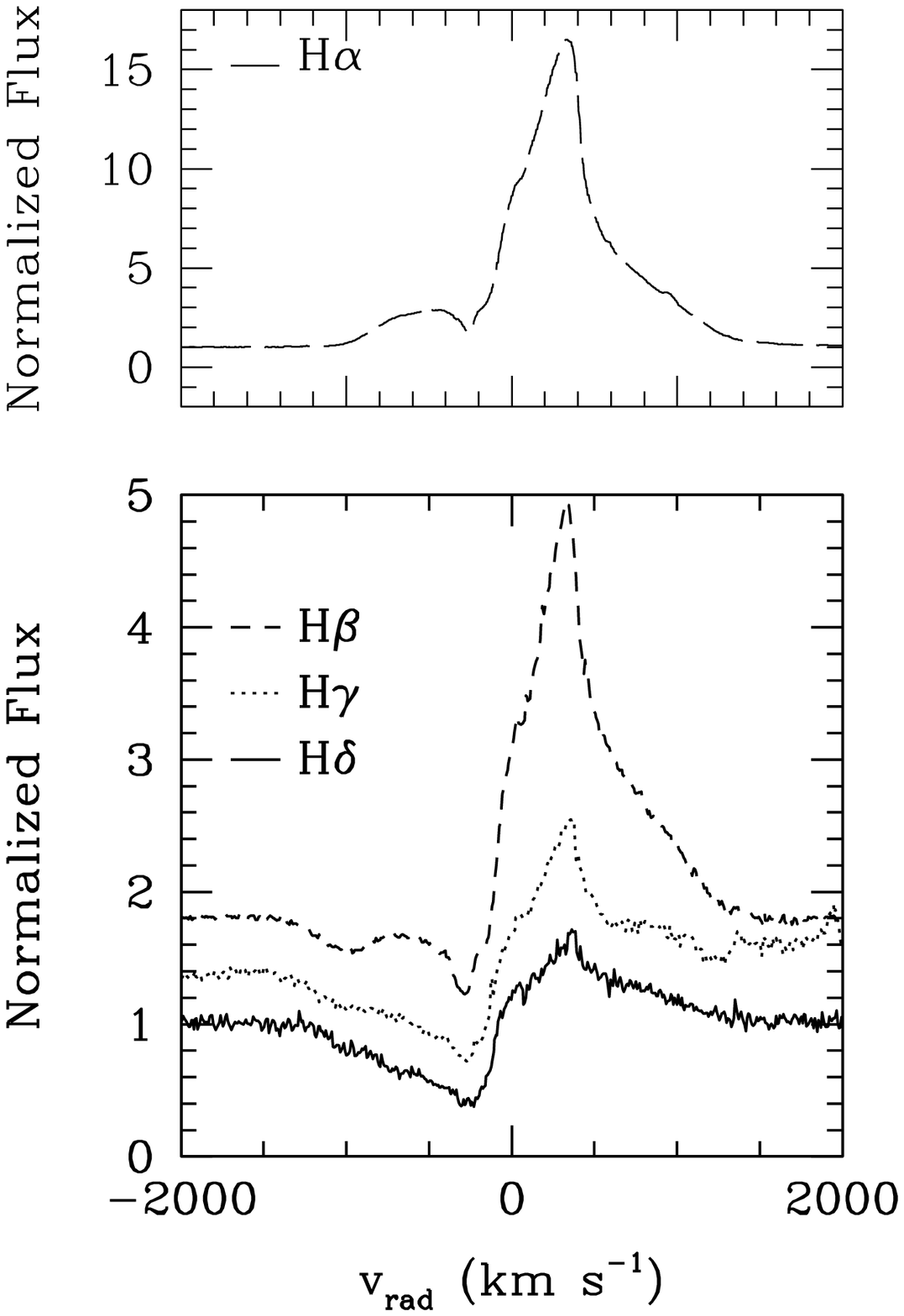}
\includegraphics[bb=30 210 580 700, clip, width=4.3cm]{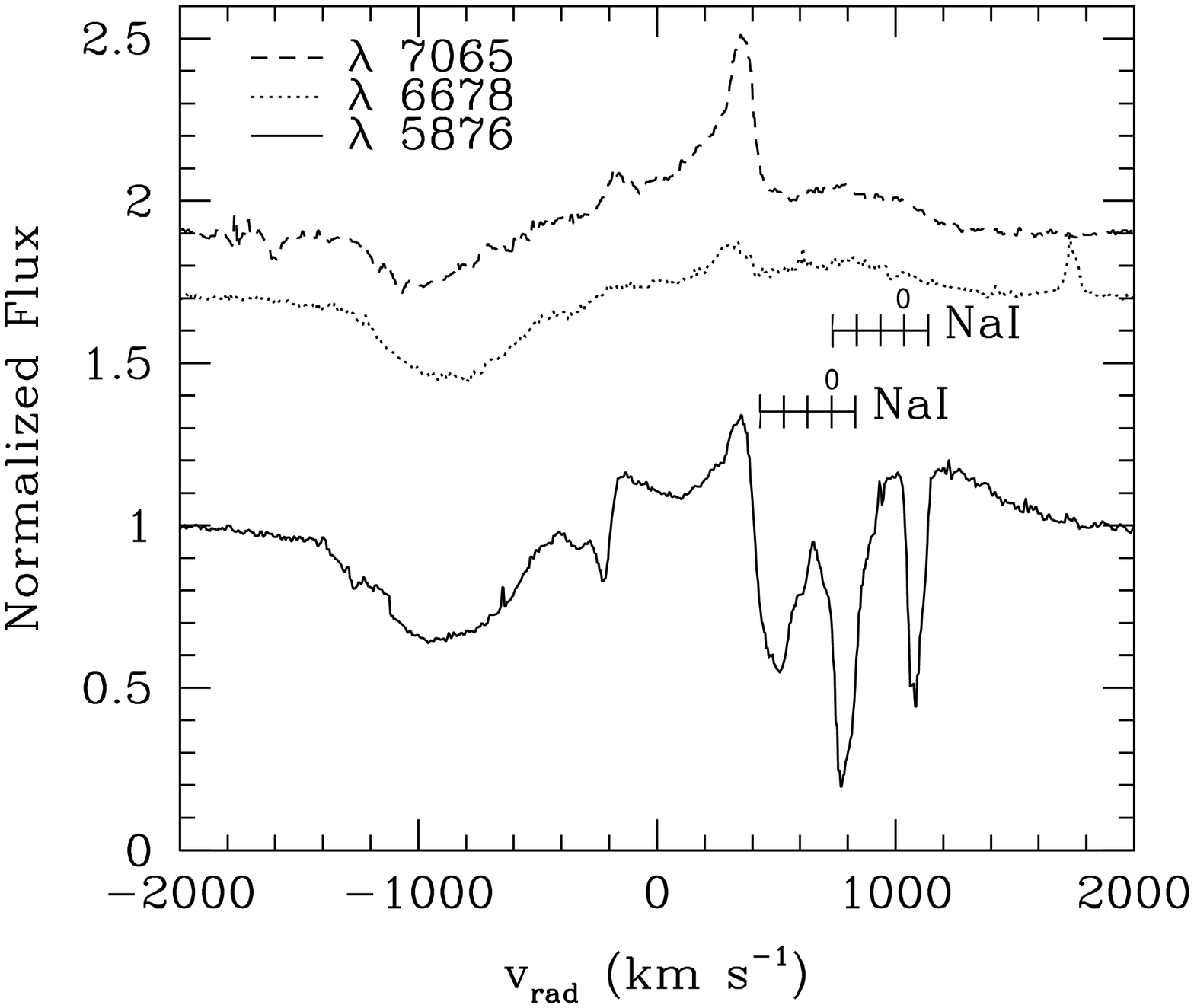}
\includegraphics[bb=30 210 580 700, clip, width=4.3cm]{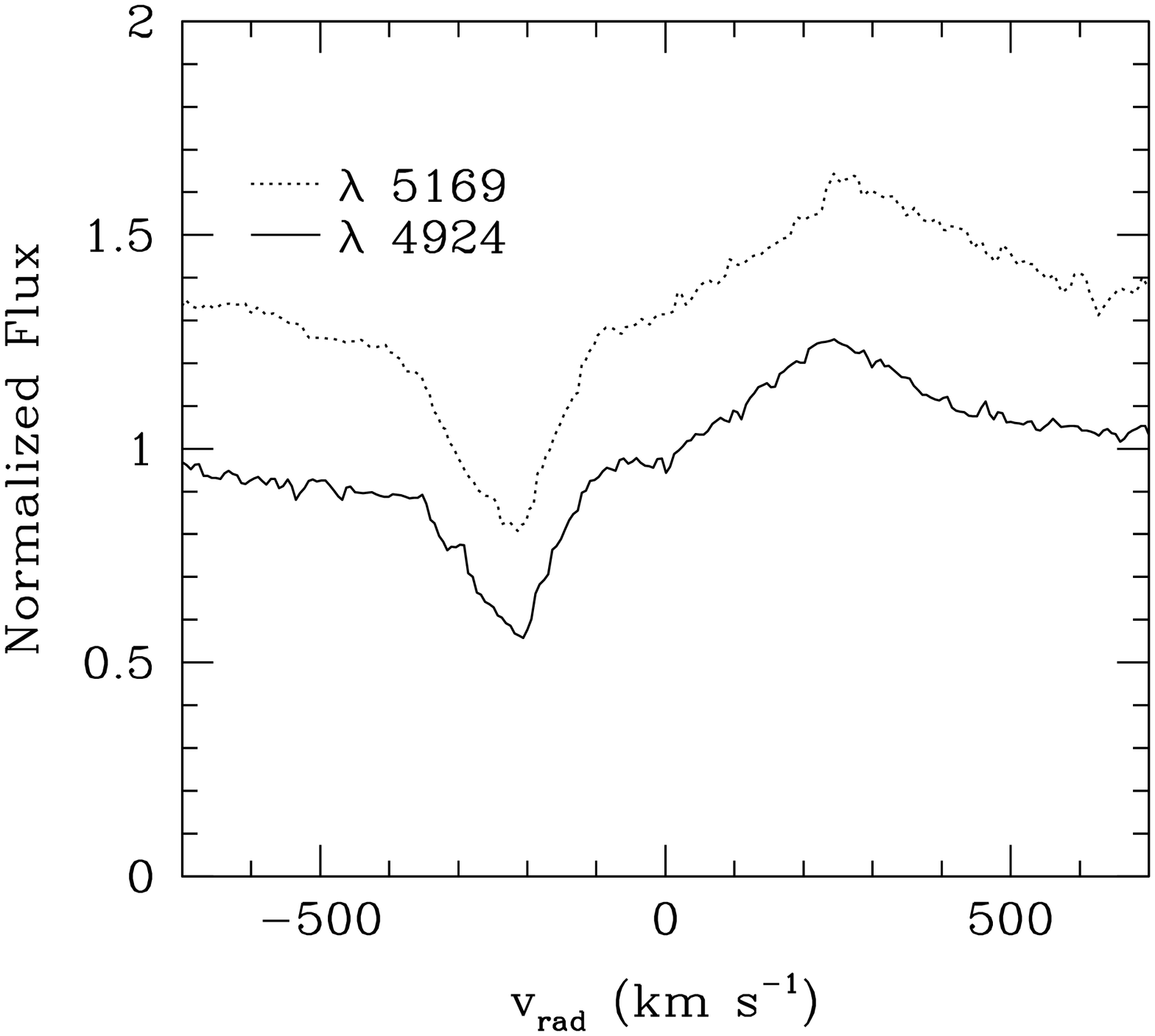}
\includegraphics[bb=30 210 580 700, clip, width=4.3cm]{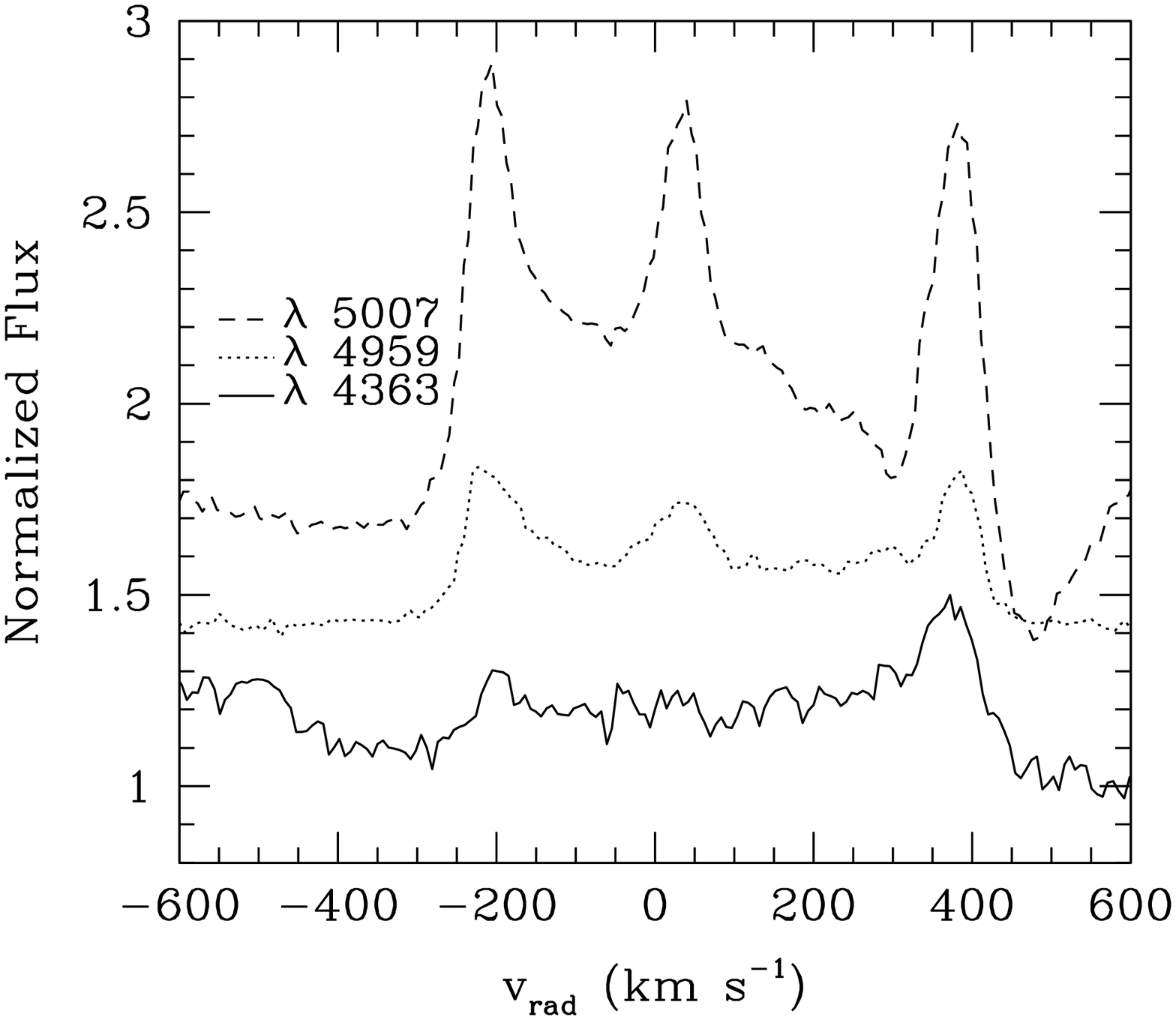}
\caption{\label{Balmer} Profiles of the H{\sc i} Balmer lines, of the most important He\,{\sc i} and  Fe\,{\sc ii} lines and of the [O{\sc iii}] lines in the March 2002 EMMI spectrum of \hen\ as a function of heliocentric radial velocity. Note that the red wing of He\,{\sc i} $\lambda$\,5876 is blended with interstellar and probably also circumstellar Na\,{\sc i} absorptions. To ease the comparison, velocity scales in the rest frame of these  Na\,{\sc i} lines are also shown, with tickmarks for each 100~\kms\ interval. The different line profiles have been arbitrarily shifted. }
\end{center}
\end{figure*}

\begin{figure}
\begin{center}
\includegraphics[width=6cm]{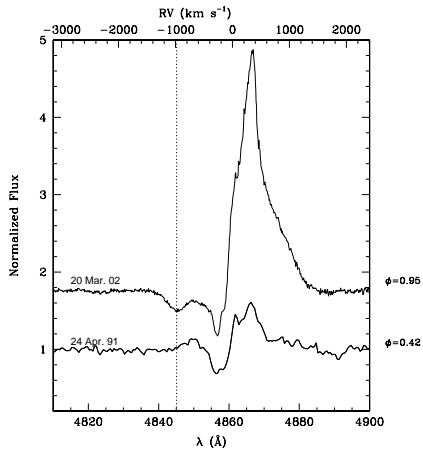}
\caption{\label{montage} Extreme spectral variations of the \hb\ line of \hen. Phases derived from the ephemeris of Sect. 3 are mentioned. A dotted line indicates the position of the broad absorption component at $-$980~\kms. Note that this component in the \hb\ line is detectable only when the emission component is very strong. }
\end{center}
\end{figure}

The hydrogen Balmer lines consist of a rather sharp emission peak at a velocity of $\sim 305$ -- $345$\,km\,s$^{-1}$ and an absorption component at $(-280 \pm 10)$\,km\,s$^{-1}$ (Fig.\,\ref{Balmer}). As was already suspected by \citet{henize}, a broader and shallower secondary absorption also appears at $-980$\,km\,s$^{-1}$ in the \ha, H$\beta$ and H$\gamma$ lines (Figs.\,\ref{Balmer} and \ref{montage}). This component is most probably also present in the H$\delta$ line, but could not be clearly disentangled from the narrow deeper component.

On the other hand, the most prominent He\,{\sc i} lines, shown in Fig.\,\ref{Balmer}, display a rather complicated profile dominated by a narrow emission peak at $\sim 325$ -- $370$\,km\,s$^{-1}$. Again, a broad absorption feature is present: it extends over about 800\,km\,s$^{-1}$ and is centered on a velocity of $\sim (-900 \pm 40)$\,km\,s$^{-1}$. At least in the case of He\,{\sc i} $\lambda$\,5876 a sharp absorption is also found at $-220$\,km\,s$^{-1}$.  

The profiles of the Fe\,{\sc ii} lines (e.g. Fe\,{\sc ii}\,$\lambda\lambda$ 4924 and 5169 presented in Fig.\,\ref{Balmer}) differ from those observed for hydrogen and helium. These iron lines display P Cygni profiles with a roughly triangular emission at $\sim 270$\,km\,s$^{-1}$ and a somewhat detached absorption component at $-230$\,km\,s$^{-1}$.

Note that an additional Na\,{\sc i} absorption  appears besides the two narrow interstellar Na\,{\sc i} features (with velocities of $50 \pm 5$\,km\,s$^{-1}$, Fig.\,\ref{Balmer}). This component displays a velocity of $-225$\,km\,s$^{-1}$, again quite close to the radial velocity found for many of the sharp absorption features seen in the H\,{\sc i}, He\,{\sc i} and Fe\,{\sc ii} P~Cygni profiles. A similar line can be spotted in the spectrum of the extreme P Cygni supergiant HDE\,316285 \citep{hillier}.

The most peculiar feature of the spectrum is the presence of the [O\,{\sc iii}] $\lambda\lambda$\,4363, 4959, 5007 forbidden lines with unusual profiles (Fig.\,\ref{Balmer}). The lines are broad, going from $\sim -200$ to $+375$\,km\,s$^{-1}$ with a saddle or flat-topped shape with three sharp components superimposed. While two of these narrow features are found at the blue and red edges of the profile, the third, slightly weaker, component has a radial velocity of $+33$\,km\,s$^{-1}$ and is therefore offset by $\sim -55$\,km\,s$^{-1}$ from the centre of the flat topped structure. In fact, this $+33$\,km\,s$^{-1}$ feature is associated with the Vela SNR\footnote{The brightest component recorded in the high-resolution CAT data displays a constant velocity of $\sim$30~\kms\ along the whole slit. This feature is also detected in the EMMI observations of the \nii, \sii, and \oiii\ lines not only at the position of the star but also next to it, suggesting that this component belongs to the Vela SNR.} which is expected to be in the foreground. The shape of the broad component and the two most extreme narrow components suggest a disk or an envelope expanding at a velocity of $\sim 300$\,km\,s$^{-1}$. Note that the large intensity of the \oiii\,\l\,4363 line suggests a high density for this feature. It should be stressed that the morphology of He\,{\sc i} $\lambda\lambda$\,5876 and 7065 between $-200$ and $+400$\,km\,s$^{-1}$ is quite reminiscent of that of the [O\,{\sc iii}] lines. The 1996 CAT data do not show the presence of these $-$200~\kms\ and +375~\kms\ components in \ha\ or \nii: both features thus display strengths similar to the foreground component in \oiii\ but are much weaker in \nii, suggesting that the N/O ratio is not unusually high. 

Since several of our spectroscopic observations have overlapping wavelength domains, we also investigated the spectral variability of \hen. Calculating the time variance spectrum (TVS, \citealt{FGB}) of our spectroscopic time series, strong variability was identified for the Balmer lines, but significant changes were also detected for the He\,{\sc i} $\lambda$\,4471, Mg\,{\sc ii} $\lambda$\,4481, Ca\,{\sc ii} H and K  as well as Fe\,{\sc ii} lines. This variability is not related to a systematic shift in radial velocity but rather consists of line profile variations, especially in the P Cygni profiles of the Balmer lines (Fig.\,\ref{montage}). These changes mainly affect the strength of the emission component and of the broad high velocity ($-980$\,km\,s$^{-1}$) absorption component, that both reach their maximum level at minimum light. The opposite situation (weak emission $and$ absorption) occurs near maximum light, suggesting dilution by additional light. However, there must be an additional cause for this variability since the equivalent widths variations are larger than the simultaneous continuum flux variations. For instance, the equivalent width of H$\gamma$ changes by a factor 4 whereas the simultaneous flux variations are only of the order of 60\%.

\section{The nature of \hen}

Our observing campaign unveiled many aspects of \hen. First, our spectra reveal spatially unresolved \oiii\ emission with  velocity components separated by up to 600 km/s. This compact emission region appears denser than the Vela SNR and shows no obvious chemical enrichment. It might be associated to an accretion disk or to material ejected by the star.

On the other hand, the optical flux from \hen\ is not constant: after a plateau, the brightness periodically undergoes a sharp, deep minimum followed by a slightly broader maximum. This behaviour could only be explained by a stable clock either due to rotation or binarity, but the shape of the lightcurve is not typical of eclipsing binaries, and requires a more complex geometry of the system (e.g. hot spots, non-alignment of the rotation and magnetic axes,...). If we envisage \hen\ as a symbiotic system, these photometric variations could reflect the occultations of the accretion disk and the central compact object by a low-mass companion or could be linked to a beating phenomenon between the orbital period and the precession period of the disk. However, these scenarios can probably be discarded since we do not find any prominent spectral features, neither in the visible domain nor in the near-infrared, attributable to a cool star companion (e.g.\ TiO bands). This lack of molecular bands rather indicates \hen\ to be a hot, early-type star.

The presence of diffuse interstellar bands (DIBs; Fig.\,\ref{echelle}) indicates that the star is rather strongly reddened and therefore located at a large distance. The interstellar extinction of \hen\ can be evaluated through the analysis of the EWs of certain DIBs \citep{HER,CR}. Using our EMMI observations and the EW data of \citet{HER} for the DIBs situated at 5797, 6195, 6269 and 6283~\AA, we found an extinction $E(B$--$V)\sim1.30$ with a scatter of 0.03~mag\footnote{Although the DIBs diagnostics are very consistent with each other, there might still be a larger, systematic error on the extinction value.}. This reddening value is indeed rather large, but an additional, circumstellar absorption may also exist. Such a large extinction can explain the lack of strong X-ray emission: an archival PSPCB observation (\ros, rp180293, 2.1~ks) only leads to an upper limit of 2.2$\times10^{-3}$~cts~s$^{-1}$ on its count rate. This non-detection would be puzzling for a nearby symbiotic binary or an X-ray binary but is not totally surprising in the case of a distant, thus strongly extinguished, early-type star. However, we may also note that \hen\ cannot be an ultra-luminous star. With a galactic longitude of $\sim270^{\circ}$, the star must be closer than $\sim$10~kpc to remain within the Galaxy. Actually, if we consider the stellar systemic velocity to be equal to the average velocity of the two extreme \oiii\ components, we find that $v_{\rm LSR}\sim76$~\kms, suggesting a distance of 8$-$9 kpc in the case of a flat galactic rotation curve \citep{mof, fic}. The above estimates of distance and interstellar reddening imply that the absolute $V$ magnitude of \hen\ should be between $-$5.6 and $-$4.7. Therefore, \hen\ cannot be a Luminous Blue Variable (LBV), although it shares several similarities with this type of stars. 

\begin{figure}
\begin{center}
\includegraphics[bb= 44 318 470 530, clip, width=7cm]{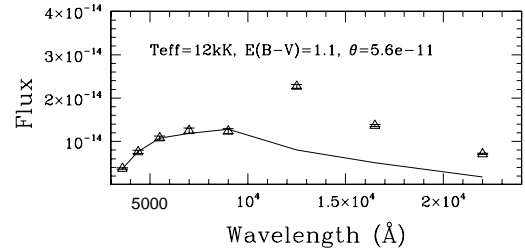}
\caption{\label{fig: armagh} The observed SED compared to a simple atmosphere model. The parameters of the best-fit model (effective temperature, reddening and angular diameter $\theta$) are noted. Note that a fit with the reddening fixed to $E(B-V)$=1.3 gives similar results and that the near-IR data were not taken into account in the fits. }
\end{center}
\end{figure}

The observed $UBVRI$ magnitudes and the estimated reddening are compatible with the intrinsic colors of B stars and giant, early B stars actually have absolute magnitudes comparable to that derived above. A fit of the apparent spectral energy distribution (SED) by a simple atmosphere model\footnote{ The software, developed at the Armagh Observatory by Simon Jeffery, is available at http://star.arm.ac.uk/$\sim$csj/software\_store } (LTE, static and plane-parallel, composition fixed to 90\%H+10\%He) also favors physical parameters similar to those of a distant B star (Fig. \ref{fig: armagh}).

Finally, \hen\ also presents peculiar IR colors. This object has been proposed to be the counterpart of the IRAS\,08464$-$4554 point source by \citet{the} but the IRAS source is rather far away (more than 30\arcsec\ from \hen) and another study rejected the identification: \citet{allen} found the IR emission of \hen\ to be a pure stellar continuum\footnote{Note however that their rather large $HK$ luminosities, $H=8.21$~mag and $K=7.93$~mag, and their proposal of an alternative candidate with more realistic magnitudes suggest a possible misidentification in their case.}. On the other hand, the 2MASS All-Sky Survey reports $J=10.44\pm0.02$, $H=9.86\pm0.02$ and $K_s=9.35\pm0.02$ for the star. Compared to ``normal" colors (Fig. \ref{fig: armagh}), \hen\ thus apparently displays a slight IR excess, possibly due  to hot dust in the compact nebula we have found, although it is by no means comparable to those of peculiar stars like $\eta$ Carinae and NaSt1 or B[e] stars \citep{CR}. In fact, the colors of \hen\ are more similar to those of normal and Wolf-Rayet stars reported by \citet[ their figure 12]{CR}. 

Considering the above evidence, our data then indicate that \hen\ probably belongs to the group of so-called {\em iron stars} that \cite{WF} introduced in their atlas of peculiar spectra. All objects of this category have prominent Fe\,{\sc ii} lines in their spectra, but they have quite varied physical natures. Some of the iron stars have been classified as B[e] stars, i.e.\ as B-type stars with forbidden emission lines in their optical spectra \citep{JPS}. \cite{Lamers} emphasized that the group of B[e] stars consists of at least five categories: supergiants (sgB[e] stars) related to the LBV phenomenon, Herbig AeB[e] pre-main sequence stars (HAeB[e] stars), compact planetary nebulae (cPNB[e] stars), symbiotic stars (SymB[e] stars) and unclassified objects (unclB[e] stars). The defining characteristics of the B[e] phenomenon are (1) strong Balmer emission lines, (2) low excitation permitted emission lines of low ionization metals, (3) forbidden emission lines of [Fe\,{\sc ii}] and [O\,{\sc i}] in the optical spectrum and (4) a strong near or mid-infrared excess due to circumstellar dust. Whilst the first two criteria are certainly met by the spectrum of \hen, we did not find strong [Fe\,{\sc ii}] nor [O\,{\sc i}] emissions nor any large IR excess. However, as pointed out above, the spectrum of \hen\ does contain high ionization forbidden emission lines  (e.g. [O\,{\sc iii}]), as are often seen for cPNB[e] stars. It is true that very young PNs can display irregular, short-term luminosity changes as well as smooth, monotonic, longer-term variations \citep{ark}, but these changes are never strictly recurrent if the star is single. This is also the case of Herbig AeBe or AeB[e] stars that display rather irregular light variations \citep{WW}. Furthermore,  the presence of P Cygni profiles in the spectrum of \hen\ suggests the presence of an outflow rather than of infalling material (that is one of the defining characteristics of the HAeB[e] category, \citealt{Lamers}). In addition, the lack of a low-mass companion further discards the SymB[e] scenario and the putative luminosity ($M_V\sim-5.1$ on average) is difficult to reconcile with a very bright sgB[e] star. In summary, though \hen\ shares some characteristics of the B[e] phenomenon, especially cPNB[e] and unclB[e], it does not meet all the criteria of this class of objects. 

With the currently available set of data, the most likely explanation for the nature of \hen\ appears to be an extreme hot star maybe belonging to a binary system and surrounded by expanding ejecta. Once disentangled from the foreground Vela SNR, a detailed study of the larger-scale nebulosities that we detect on narrow-band images and possibly associated to the star may shed more light on the mass-loss history and nature of \hen.

\section{Conclusions}

Our study shed new light on the poorly known peculiar system \hen. In our data, this intriguing object displays recurrent photometric variations with a peak-to-peak amplitude of 0.65~mag and a period of 16.09$\pm$0.01~days. Its spectrum presents many P~Cygni profiles (H\,{\sc{i}}, \he, \feii,...) and some forbidden lines like \oiii. The Balmer line profiles vary along with the photometry. The \oiii\ profile is very peculiar and contains three sub-peaks: one is associated with the foreground nebula, but the other two, separated by $\sim$600~\kms, indicate the presence of an accretion disk or of expanding ejecta close to the star. 

The actual nature of \hen\ is still difficult to ascertain with our limited sample of observations, but most existing evidence points towards a moderately bright and distant object ($d\sim8-9$~kpc and $M_V\sim-5.1$) having probably undergone a mass ejection event and maybe belonging to a binary system.

\section*{Acknowledgments}

We thank O. Hainaut and I. Saviane for their help in collecting some of the data and Simon Jeffery for his advices and guidelines for using his atmosphere code. We also acknowledge support from the FNRS (Belgium) and from the {\sc PRODEX XMM} and Integral contracts. This research is also partly supported by contracts P5/36 ``PAI'' (Belspo). This publication makes use of Simbad, the ADS abstract service and the data products from the 2MASS.

\label{lastpage}

\begin{thebibliography}{99}
\bibitem[\protect\citeauthoryear{Allen \& Glass}{1975}]{allen} Allen, D.A., Glass, I.S. 1975, MNRAS, 170, 579 
\bibitem[\protect\citeauthoryear{Arkhipova et al.}{2001}]{ark} Arkhipova, V.P., Ikonnikova, N.P., Noskova, R.I., Komissarova, G.V., Klochkova, V.G., Esipov, V.F. 2001, Astr. Letters, 27, 719
\bibitem[\protect\citeauthoryear{Crowther \& Smith}{1999}]{CR} Crowther, P.A., Smith, L.J.\ 1999, MNRAS, 308, 82
\bibitem[\protect\citeauthoryear{de Winter et al.}{2001}]{dew} de Winter, D., van den Ancker, M.E., Maira, A., et al. 2001, A\&A, 380, 609
\bibitem[\protect\citeauthoryear{Fich, Blitz, \& Stark}{Fich et al.}{1989}]{fic} Fich, M., Blitz, L., Stark, A.A. 1989, ApJ, 342, 272
\bibitem[\protect\citeauthoryear{Fullerton et al.}{1996}]{FGB} Fullerton, A.W., Gies, D.R., Bolton, C.T.\ 1996, ApJS, 103, 475 
\bibitem[\protect\citeauthoryear{Gosset et al.}{2001}]{GOS} Gosset, E., Royer, P., Rauw, G., Manfroid, J., Vreux, J.-M.\ 2001, MNRAS, 327, 435
\bibitem[\protect\citeauthoryear{Heck, Manfroid \& Mersch}{Heck et al.}{1985}]{HMM} Heck, A., Manfroid, J., Mersch, G. 1985, A\&AS, 59, 63
\bibitem[\protect\citeauthoryear{Henize}{1976}]{henize} Henize, K.G.\ 1976, ApJS, 30, 491
\bibitem[\protect\citeauthoryear{Herbig}{1995}]{HER} Herbig, G.H.\ 1995, ARA\&A, 33, 19
\bibitem[\protect\citeauthoryear{Hillier et al.}{1998}]{hillier} Hillier, D.J., Crowther, P.A., Najarro, F., Fullerton, A.W.\ 1998, A\&A, 340, 483
\bibitem[\protect\citeauthoryear{Lafler \& Kinman}{1965}]{LK} Lafler, J., Kinman, T.D. 1965, ApJS, 11, 216
\bibitem[\protect\citeauthoryear{Lamers et al.}{1998}]{Lamers} Lamers, H.J.G.L.M., Zickgraf, F.-J., de Winter, D., Houziaux, L., Zorec, J.\ 1998, A\&A, 340, 117
\bibitem[\protect\citeauthoryear{Moffat et al.}{1998}]{mof} Moffat, A.F.J., Marchenko, S.V., Seggewiss, W., et al. 1998, A\&A, 331, 949
\bibitem[\protect\citeauthoryear{Pojmanski}{2002}]{poj} Pojmanski, G. 2002, Acta Astronomica, 52, 397
\bibitem[\protect\citeauthoryear{Roberts}{1962}]{roberts} Roberts, M.S.\ 1962, AJ, 67, 79
\bibitem[\protect\citeauthoryear{Sanduleak \& Stephenson}{1973}]{sand} Sanduleak, N., Stephenson, C.B. 1973, ApJ, 185, 899
\bibitem[\protect\citeauthoryear{Smith}{1968}]{smi} Smith, L.F.\ 1968, MNRAS, 138, 109
\bibitem[\protect\citeauthoryear{Stetson}{1987}]{STE} Stetson, P.B. 1987, PASP, 99, 191
\bibitem[\protect\citeauthoryear{Swings}{1976}]{JPS} Swings, J.-P.\ 1976, in {\it Be and Shell Stars}, IAU Symp. 70, ed.\ A.\ Slettebak, Reidel, Dordrecht, p219
\bibitem[\protect\citeauthoryear{Th\'e, de Winter \& P\'erez}{Th\'e et al.}{1994}]{the} Th\'e, P.S., de Winter, D., P\'erez, M.R.\ 1994, A\&AS, 104, 315
\bibitem[\protect\citeauthoryear{Velghe}{1957}]{vel} Velghe, A.G. 1957, ApJ, 126, 302
\bibitem[\protect\citeauthoryear{Walborn \& Fitzpatrick}{2000}]{WF} Walborn, N.R., Fitzpatrick, E.L.\ 2000, PASP, 112, 50
\bibitem[\protect\citeauthoryear{Waters \& Waelkens}{1998}]{WW} Waters, L.B.F.M., Waelkens, C.\ 1998, ARA\&A, 36, 233
\end{thebibliography}
\end{document}